\documentclass[prl,superscriptaddress,twocolumn,floatfix,showpacs,letterpaper]{revtex4}
\usepackage{amsmath}
\usepackage{graphicx}
\usepackage{times}
\usepackage{hyperref}
\begin{document}

\title{New Phases of Water Ice Predicted at Megabar Pressures}

\author{Burkhard Militzer} 
\affiliation{Department of Earth and Planetary Science}
\affiliation{Department of Astronomy, University of California, Berkeley}
\author{Hugh F. Wilson} 
\affiliation{Department of Earth and Planetary Science}

\begin{abstract}

  Based on density functional calculations we predict water ice to
  attain two new crystal structures with {\it Pbca} and {\it Cmcm}
  symmetry at 7.6 and 15.5 Mbar, respectively. The known high pressure
  ice phases VII, VIII, X, and {\it Pbcm} as well as the {\it Pbca}
  phase are all insulating and composed of two interpenetrating
  hydrogen bonded networks, but the {\it Cmcm} structure is metallic
  and consists of corrugated sheets of H and O atoms. The H atoms are
  squeezed into octahedral positions between next-nearest O atoms
  while they occupy tetrahedral positions between nearest O atoms in
  the ice X, {\it Pbcm}, and {\it Pbca} phases.

\end{abstract} 

\maketitle

Water ice is one of the most prevalent substances in the solar system,
with the majority of it existing at high pressures in the interiors of
giant planets~\cite{hubbard_planets}. Uranus and Neptune are assumed
to consist largely of a mixture of water, ammonia and methane ices at
pressures up to 8 Mbar~\cite{guillot-science-99}, while Jupiter and
Saturn almost certainly have large dense cores consisting of
differentiated rock and ice components at pressures on the order of 10
Mbar in Saturn~\cite{guillot-science-99} and 39$-$64 Mbar in
Jupiter~\cite{MHVTB}. The behavior of water ice under these extreme
conditions is not yet well understood because static diamond anvil
cell experiments have not yet reached beyond 2.1
megabars~\cite{hemley-ice-1987,Goncharov1996,Loubeyre1999,Goncharov2009}. Shock wave
experiments~\cite{lee2006} have reached higher pressures but they also
heat the sample significantly so that it melts for the highest
pressures.  However, dynamic ramp compression
techniques~\cite{Davis2005,Dolan2009} are expected to reach high
pressures at comparatively low temperatures, where so far only
theoretical methods have predicted the state of water ice.

The known phase diagram of water is extremely rich, with at least
fifteen forms of solid ice observed
experimentally~\cite{salzmann-prl-09} and one high-pressure phase
predicted theoretically~\cite{benoit-prl-96}. The high-pressure region
of the phase diagram is comparatively simple. The molecular ice VIII
phase, which forms at low temperatures for pressures above $\sim$15
kbar, consists of a {\it bcc} array of oxygen atoms with an ordered
arrangement of hydrogen atoms arranged along the tetrahedral
directions, bonded with a short covalent bond to one oxygen and a
longer hydrogen bond to the other. Increasing pressure to 0.7 Mbar
results in the symmetrization of these bonds~\cite{polian-prl-84},
with the distinction between covalent and hydrogen bonds being lost as
the hydrogen occupies the midpoint between the two oxygens. This is
referred to as phase X, the highest-pressure phase that has been
observed experimentally.  In 1996 Benoit \emph{et
  al.}~\cite{benoit-prl-96} used density functional theory (DFT) to
predict a higher-pressure phase of ice with {\it Pbcm} symmetry and
twelve atoms per unit cell to become stable at approximately 3~Mbar.
It was recently shown within DFT that this phase forms by a dynamic
instability in the ice X lattice~\cite{caracas-prl-08,marques2009}.

At temperatures in excess of approximately 2000~K, high-pressure ice
transitions to a superionic
phase~\cite{cavazzoni,goncharov2005,mattson2006} in which the hydrogen
atoms become mobile while the oxygen atoms do not, while at higher
temperatures still the oxygen atoms also become mobile and the entire
structure melts~\cite{schwegler-PNAS-2008}. In the recent work by
French \emph{et al.}~\cite{french-prb-09} it was shown that the {\it
  bcc} structure of the oxygen lattice in the superionic phase appears
to be maintained up to very high pressures.  This work also indicated
that at densities below approximately 5 g$\,$cm$^{-3}$, corresponding
to pressures around 6 Mbar, the hydrogen atoms are found to be
strongly associated with the four tetrahedral sites surrounding each
oxygen as in ice X, but at higher pressures the hydrogen atoms show an
increasing preference for the six octahedral sites surrounding around
each oxygen. It is thus natural to ask whether the occupation of these
sites may lead to the formation of a novel crystalline phase of ice
that would be more stable than the {\it Pbcm} phase at high pressure.

To investigate the plausibility of the occupation of the octahedral
sites at low temperature we performed a molecular dynamics simulation
in which a superionic water sample in a 48-atom cell at a pressure of
approximately 29 Mbar was quenched from a temperature of 5000~K to
zero temperature over a period of 2~ps. In this simulation, as with
all simulations described later, we used the VASP density functional
theory code~\cite{vasp1}, with pseudopotentials of the
projector-augmented wave type~\cite{PAW}, a cutoff for the expansion
of the plane wave basis set for the wavefunctions of at least 1360~eV,
and the PBE exchange-correlation functional~\cite{PBE}. The simulation
used an NVT ensemble controlled by a Nose-Hoover thermostat in which
the cell vectors were not allowed to change. In the resulting
structure, the oxygen atoms retained their {\it bcc} lattice
structure, while the hydrogen atoms all ended up close to the
octahedral lattice sites, while none was close to a tetrahedral
lattice site. This provides motivation for a more thorough
investigation of this class of crystalline structures with octahedral
hydrogen occupation.

\begin{figure}[htbl]
\includegraphics[width=0.41\textwidth]{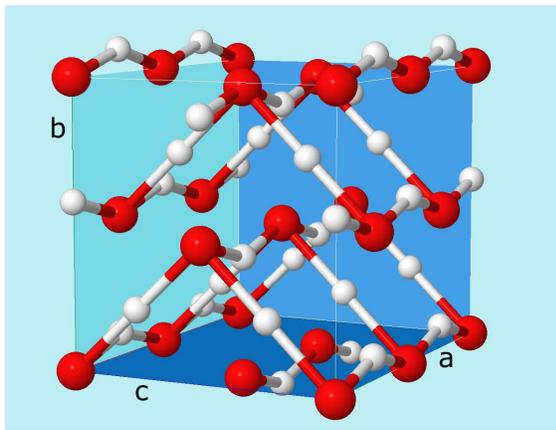}
\caption{{\it Cmcm} ice structures where the large and small spheres
  denote the oxygen and hydrogen atoms respectively. The orthorhombic
  unit cell with 12 atoms has been doubled in $a$ direction and
  shifted so that an oxygen atom is at the origin.}
\label{fig3D}
\end{figure}

The {\it bcc} oxygen lattice provides four tetrahedral sites
surrounding each oxygen and lying at the midpoint of the
nearest-neighbor pairs, and six octahedral sites lying at the midpoint
of second-nearest-neighbor pairs. This gives only one way to occupy
the tetrahedral sites with two hydrogen atoms per oxygen, but many
ways to occupy the octahedral sites with two hydrogen atoms per
oxygen. On the assumption that the stablest configuration was likely
to have reasonably high symmetry, we systematically studied each
possible configuration of hydrogen atoms for the three smallest
possible unit cells: the six-atom $1 \times 1 \times 1$ unit cell, the
twelve-atom $2 \times 1 \times 1$ unit cell and the twelve-atom
$\sqrt{2} \times \sqrt{2} \times 1$ unit cell. This resulted, after
the removal of equivalent structures, in three configurations for the
$1 \times 1 \times 1$ unit cell, six configurations for the $2 \times
1 \times 1$ unit cell and also six configurations for the $\sqrt{2}
\times \sqrt{2} \times 1$ unit cell. Larger unit cells were found to
have impractically large numbers of inequivalent configurations. For
each configuration, we performed a full geometry optimization at a
pressure of 29 Mbar.

The most stable configuration consists of a $\sqrt{2} \times \sqrt{2}
\times 1$ orthorhombic cell with 12 atoms and is shown in
Fig.~\ref{fig3D}. Its reduced coordinates given in Tab.~\ref{tab1}.
The structure has {\it Cmcm} symmetry and may also be represented in
6-atom monoclinic unit cell with the vectors $a_{\rm m}$=$(a+b)/2,
b_{\rm m}$=$(a-b)/2, c_{\rm m}$=$c$.
This structure is one of a relatively small number of configurations
which can be formed by filling two out of every three octahedral
lattice sites with hydrogen atoms in such a way that no two vacancies
are immediately adjacent. During the structural relaxation, the
hydrogen atoms that connect two oxygen atoms along the $a$ direction
move up along the $b$ direction ($y=0.25 \to 0.34$) towards the third
vacant hydrogen site.  This elongates their bonds with the nearest
oxygen atoms and introduces the kinks into O-H chains that are shown
in Fig.~\ref{fig3D}. 
The structural relaxation of the next three
lowest-enthalpy structures yielded enthalpies which are 0.51, 0.59,
and 1.1 eV per H$_2$O unit higher. 
We also doubled and quadrupled the unit cell and re-relaxed structure
starting with distorted hydrogen positions but no structure with lower
enthalpy was found.

\begin{figure}[htbl]
\includegraphics[width=0.47\textwidth]{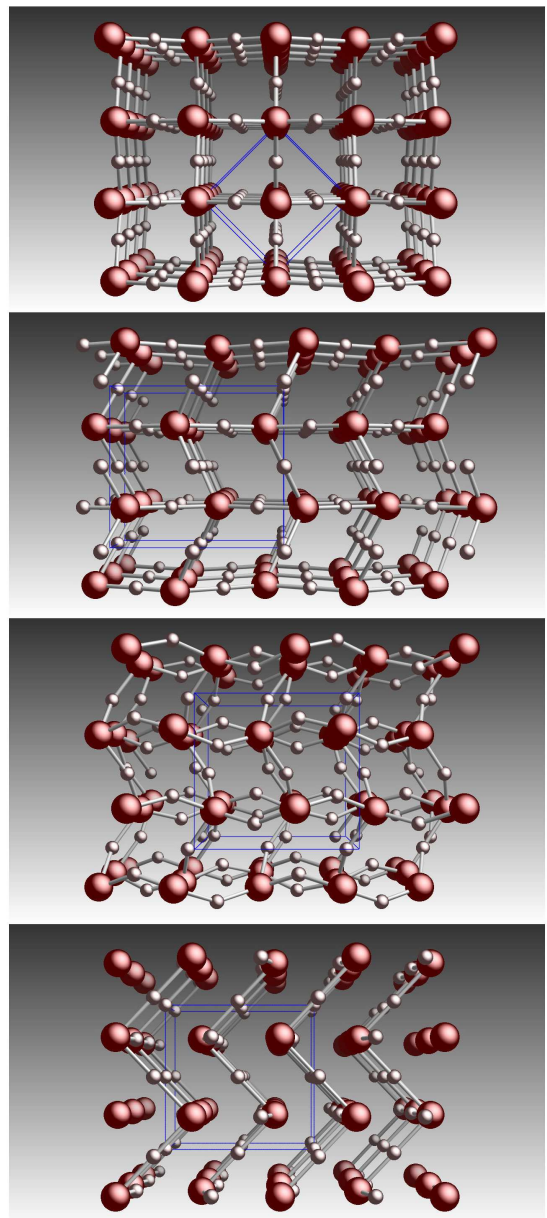}
\caption{Structures of the ice X, {\it Pbcm}, and new {\it Pbca} and
  {\it Cmcm} phases are shown from top to bottom. The large and small
  spheres denote the O and H atoms respectively, while the thin lines
  denote the unit cells. The ice X to {\it Pbcm} transition is a
  displacement of atomic layers. In {\it Pbca}, the H atoms are
  squeezed out of midpoint between nearest O atoms. In {\it Cmcm}, the
  H atoms occupy mid-points between next-nearest O atoms. }
\label{four_ices}
\end{figure}

The orthorhombic structure with {\it Cmcm} symmetry is common among
high pressure materials including phase II of molecular
hydrogen~\cite{Toledano2009}, AB and AB$_2$ compounds, and the
post-perovskite phase of ABO$_3$. A {\it Cmcm}
structure~\cite{Neaton2002} with a four-atom unit cell was also
proposed for the $\epsilon$ phase of molecular oxygen but experiments
later determined a different structure with $C2/m$
symmetry~\cite{Lundegaard}.

\begin{table}
\centering
{\footnotesize
\begin{ruledtabular}
\begin{tabular}{l c c c }
Symmetry                 & {\it Pbca} & {\it Pnma} & {\it Cmcm} \\
Pressure (Mbar)          & 10.0        & 12.0        & 20.0        \\
Density (g$\,$cm$^{-3}$) & 6.192       & 6.679       & 8.115       \\
Atoms in unit cell       & 24          & 12          & 12          \\
\hline
a (\AA) & 3.117 & 3.005 & 1.869\\
b (\AA) & 3.762 & 1.925 & 2.841\\
c (\AA) & 3.295 & 3.098 & 2.778\\
\hline
H      & $c$ $\left(\!\!\begin{array}{c} 0.8036\\   0.0015\\   0.1484\end{array}\!\!\right)$ & 
         $c$ $\left(\!\!\begin{array}{c} 0.0029\\   3/4\\      0.2330\end{array}\!\!\right)$ & 
         $a$ $\left(\!\!\begin{array}{c} 0\\        0\\        0\end{array}\!\!\right)$ \\

H      & $c$ $\left(\!\!\begin{array}{c} 0.4862\\   0.7161\\   0.5370\end{array}\!\!\right)$ & 
         $c$ $\left(\!\!\begin{array}{c} 0.2870\\   3/4\\      0.9154\end{array}\!\!\right)$ & 
         $c$ $\left(\!\!\begin{array}{c} 1/2\\      0.8498\\   1/4\end{array}\!\!\right)$ \\

O      & $c$ $\left(\!\!\begin{array}{c} 0.2592\\   0.1311\\   0.1040\end{array}\!\!\right)$ & 
         $c$ $\left(\!\!\begin{array}{c} 0.7536\\   3/4\\      0.0427\end{array}\!\!\right)$ & 
         $c$ $\left(\!\!\begin{array}{c} 1/2\\      0.7237\\   3/4\end{array}\!\!\right)$ \\
\end{tabular}
\end{ruledtabular}} 
\caption{Structural parameters of different ice phases in orthorhombic unit cells. The last three lines specify the Wyckoff 
  positions and the reduced coordinates of the atoms. The remaining positions follow from symmetry operations.}
\label{tab1}
\end{table}

\begin{figure}[htbl]
\includegraphics[width=0.41\textwidth]{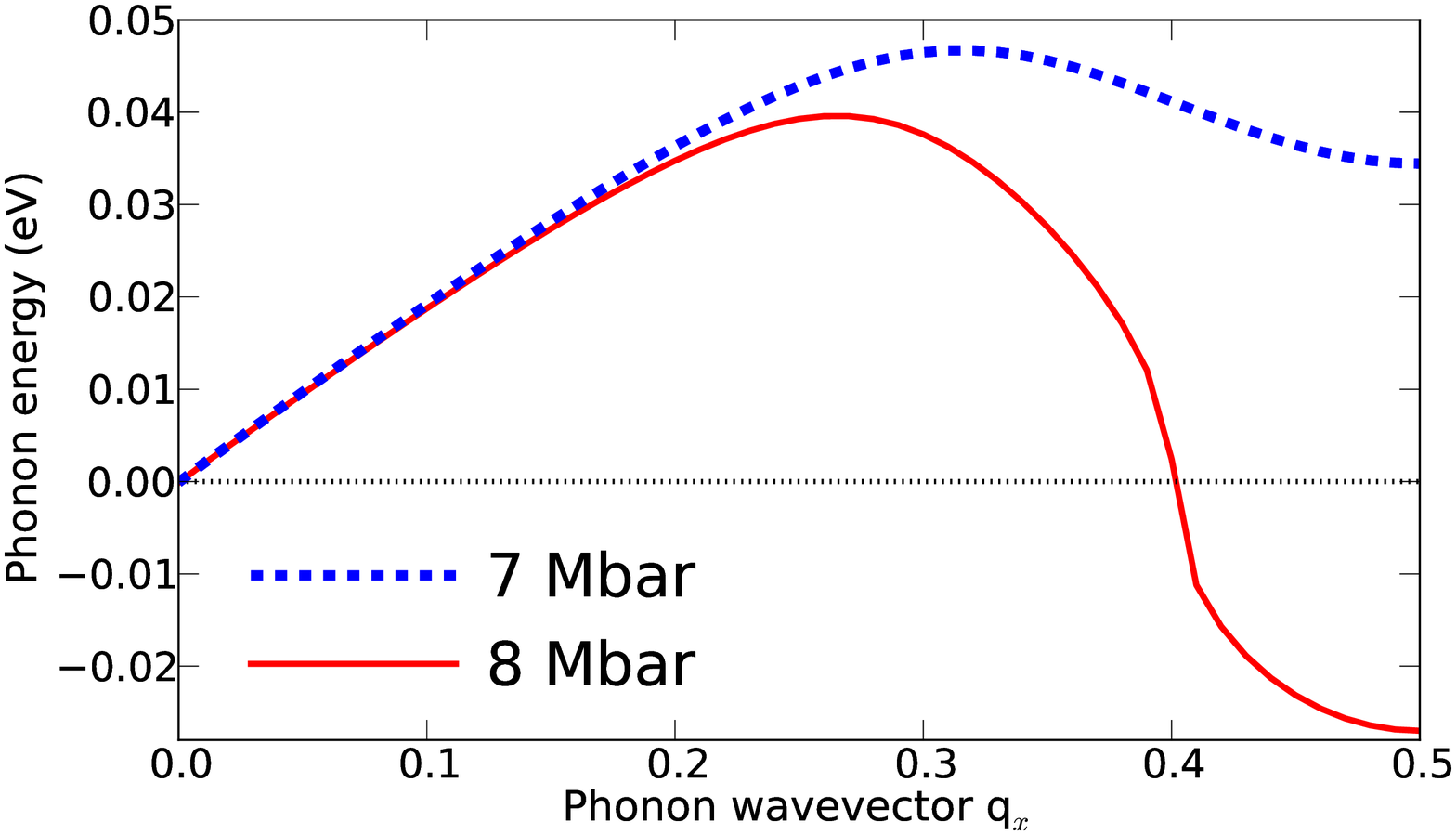}
\caption{Dispersion curve of the {\it Pbcm} phonon band with the
  lowest energy along the $a$ direction, exhibiting an instability at 7.6 Mbar.}
\label{phonons}
\end{figure}

Prompted by the referee, we also performed lattice dynamics
calculations with Abinit~\cite{Abinit2009} and VASP codes. This revealed a dynamic
instability in the {\it Pbcm} structure. Fig.~\ref{phonons}
illustrates that the mode $(1/2,0,0)$ becomes unstable at 7.6 Mbar.
The structural relaxation in a $2 \times 1 \times 1$ unit cell with 24
atoms gave rise to a new intermediate structure of {\it Pbca} symmetry
(Tab.~\ref{tab1}) that precedes the transition to the {\it Cmcm} phase in
pressure. The resulting sequence is shown in Fig.~\ref{four_ices}. In
the {\it Pbca} structure, the hydrogen atoms are squeezed out of the
midpoint between two nearest oxygen atoms.  However, they still reside
near their tetrahedral sites. The distortion of the H positions occurs
in alternating directions in the two hydrogen bonded networks, which
is accommodated by the unit cell doubling in $a$ direction.

\begin{figure}[htbl]
\includegraphics[width=0.46\textwidth]{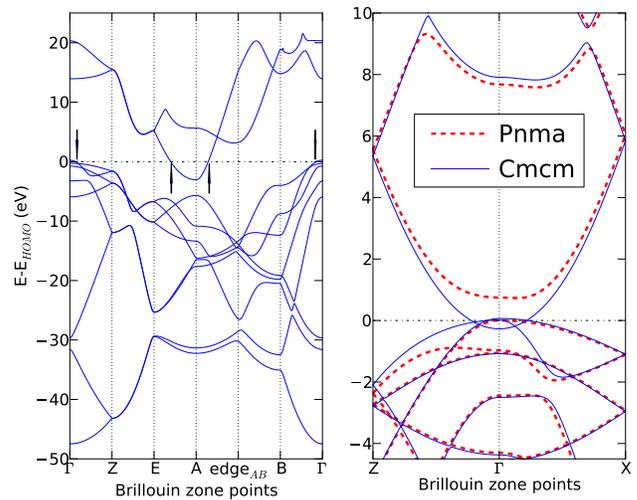}
\caption{Left: Electronic band structure for the {\it Cmcm} phase in a
  monoclinic unit cell at 22 Mbar.  The arrows indicate where bands
  cross the Fermi level underlining the metallic nature of this phase.
  Right: Comparison of {\it Pnma} and {\it Cmcm} band structures at 12.5 Mbar in
  the orthorhombic unit cell illustrate the Peierls instability.}
\label{bands}
\end{figure}


The {\it Cmcm} structure is distinct from the other high-pressure ice
phases in several ways. While ices VII, VIII and X as well as the {\it
  Pbcm} and {\it Pbca} structures all consist of two interpenetrating
hydrogen bonded networks, the {\it Cmcm} structure consists of
corrugated sheets of H and O atoms
(Figs.~\ref{fig3D}~\&~\ref{four_ices}). While in lower-pressure
structures the hydrogen atoms occupy sites between nearest pairs of
oxygen atoms, in the \emph{Cmcm} structure they are squeezed into
sites between second-nearest pairs of oxygen atoms. It should be noted
that the characterization of crystalline solids in terms of chemical
bonds makes less sense at very high pressure because the structures
are increasingly dominated by the mutual repulsion of atomic cores
rather than the formation of favorable electronic bonds between atoms.
The repulsive forces that keep the atoms at their lattice sites
increase strongly with pressure, however.

Band structure calculations show that the {\it Cmcm} phase is metallic for
all pressures under consideration. Fig.~\ref{bands} shows that the
band which was the conduction band in the lower-pressure structures,
now dips below the Fermi level a the A point of the monoclinic cell
and becomes partially occupied. Similarly a valence bands becomes
unoccupied near $\Gamma$. 

However, careful structural relaxations in the pressure range of
7.6-15.5 Mbar reveal the existence of a Peierls instability that opens
a band gap (Fig.~\ref{bands}) by shifting hydrogen atom slightly away
from the mid-point between near-nearest oxygen atoms. This distortion
is small and fractional coordinates change by less than 2\%. It lowers
the enthalpy by just 7 meV per molecule. The resulting structure has
{\it Pnma} symmetry with 12 atoms in an orthorhombic unit cell.
Table~\ref{tab1} lists the structural parameters in the conventional
coordinate setting for {\it Pnma} that differs from that for {\it Cmcm}.

At 15.5 Mbar the band gap in the {\it Pnma} structure closes and the
mechanism for the Peierls instability disappears. This pressure marks
the transition from the {\it Pnma} to the {\it Cmcm} structure. Since
all lower pressure ice phases are insulating, the transition to {\it
  Cmcm} at 15.5 Mbar marks the insulator-to-metal transition in water
ice.

\begin{figure}[htbl]
\includegraphics[width=0.47\textwidth]{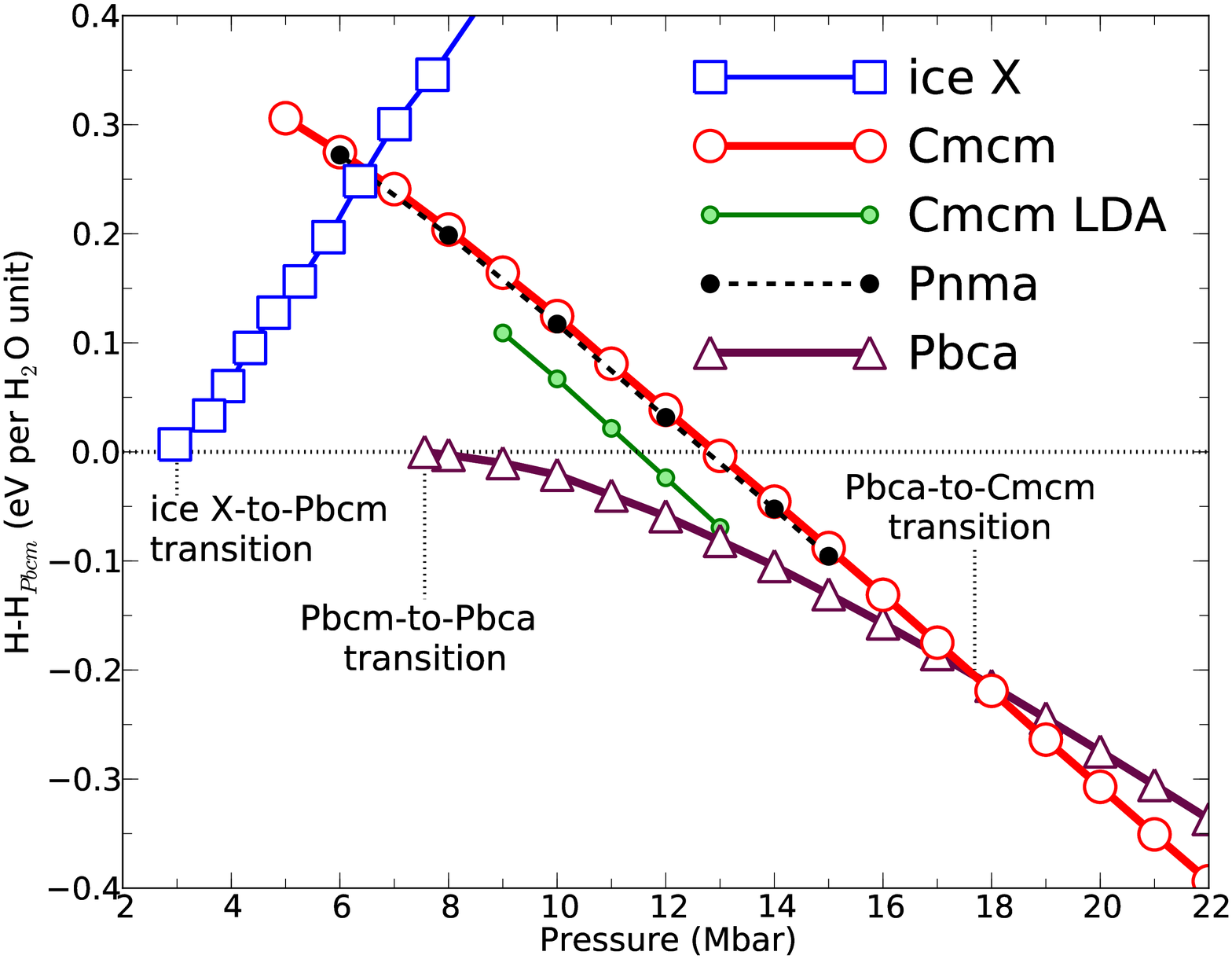}
\caption{Enthalpy differences of the ice X, {\it Pbca}, {\it Pnma}, and {\it Cmcm}
  structures from the {\it Pbcm} structure as a function of pressure.
  PBE~\cite{PBE} was used except where LDA is indicated. Results are
  for static lattices without zero point motion.}
\label{figH}
\end{figure}

We now compare the enthalpy, $H=E+PV$, of the ice X, {\it Pbcm}, {\it
  Pbca}, {\it Pnma}, and {\it Cmcm} structure after having optimized
the geometry in constant-pressure variable-cell simulations from 5 to
50 Mbar. Figure~\ref{figH} shows the enthalpy difference of the {\it
  static} lattice relative to the \emph{Pbcm} structure. The ice X
structure transforms into the \emph{Pbcm} structure for pressures
above approximately 3 Mbar, consistent with the results of Benoit
\emph{et al.}~\cite{benoit-prl-96}.  The \emph{Pbcm} structure
transforms into the \emph{Pbca} phase at 7.6 Mbar consistent with
our lattice dynamics calculations. 

According to static lattice calculations, the {\it Pbca} phase would
transform into {\it Cmcm} structure at 17.7 Mbar. To test the
importance of zero point motion, we perform lattice dynamics
calculation in the harmonic approximation with a $4 \times 4 \times 4$
$q$ point grid. The {\it Cmcm} and {\it Pnma} phases have
significantly lower zero point energy than the {\it Pbca} phase (e.g.
75 meV per molecule at 12 Mbar) because the hydrogen atoms are less
constrained at their octahedral sites between the more distant
next-nearest oxygen atoms, leading to softer phonon modes. The
resulting reduction in zero point energy means that the {\it Pbca}
structure transforms into the insulting {\it Pnma} structure at 12.5
Mbar before this structure changes to {\it Cmcm} at 15.5 Mbar.


Prompted by the referee, we also tested the validity of the PAW VASP
pseudopotentials by performing full-potential all-electron calculation
with the {\it Exciting} code~\cite{exciting}. We focused this test on
the \emph{Pbcm}-to-\emph{Cmcm} transition that occurs at 13.1 Mbar
(Fig.~\ref{figH}) within the PBE approximation. We recalculated this
transition pressure within the local density approximation (LDA) and
obtained 11.5 Mbar. Such a pressure difference is not unexpected
because LDA typically underestimates transition
pressures~\cite{Hamann96}. We then determined the energy from
all-electron calculation for the five geometries in each phase that we
obtained with LDA PAW calculations from 9 to 14.3 Mbar.  We fit the
resulting equations for state for both phases and recalculated the
transition pressure. The all-electron method shifted the resulting
transition pressure by less than 0.1 Mbar, which confirms that the PAW
pseudopotentials are sufficiently accurate for the purpose for this
study.

%
%
We have predicted that water ice will attain novel crystal structures
at megabar pressures and determined the following sequence of structural
transformations.
At 7.6 Mbar the {\it Pbcm} phase will transform into a {\it Pbca}
phase, which then changes into an insulating {\it Pnma} structure at
12.5 Mbar. At 15.5 Mbar a insulator-to-metal transition leads to a
structure with {\it Cmcm} symmetry. This last transition is expected
to greatly increase reflectivity in water ice, which will make it
easier detect with spectroscopic techniques in dynamic high pressure
experiments. Ramp compression~\cite{Davis2005,Dolan2009} and
pre-compressed~\cite{KananiLee2006} shock waves appear as most
promising techniques.


\acknowledgments{This work was supported by NASA and NSF.
  Computational resources were provided in part by NCCS, NERSC, and
  TAC. We acknowledge discussions with T. Ogitsu, R. Jeanloz, and R.
  Martin and advice on phonon calculations from S.  Stackhouse and I.
  Souza.}
\end{document}